\documentclass[12pt]{article}

\usepackage{graphicx}
\begin{document}

\begin{center}
{\bf Nonlinear electrodynamics and magnetic black holes} \\
\vspace{5mm} S. I. Kruglov
\footnote{E-mail: serguei.krouglov@utoronto.ca}
\underline{}
\vspace{3mm}

\textit{Department of Chemical and Physical Sciences, University of Toronto,\\
3359 Mississauga Road North, Mississauga, Ontario L5L 1C6, Canada} \\
\vspace{5mm}
\end{center}
\begin{abstract}
 A model of nonlinear electrodynamics with two parameters, coupled with general relativity, is  investigated. We study the magnetized black hole and obtain solutions.
The asymptotic of the metric and mass functions at $r\rightarrow\infty$ and $r\rightarrow 0$, and
corrections to the Reissner-Nordstr\"{o}m solution are found. We investigate thermodynamics of black holes and calculate the Hawking temperature and heat capacity of black holes. It is shown that there are phase transitions and at some parameters of the model black holes are stable.
\end{abstract}

\section{Introduction}

In last years models of nonlinear electrodynamics (NLED) attract attention of astrophysicists.
Some models of NLED can explain inflation of the universe by NLED coupled to general relativity \cite{Garcia}-\cite{Kruglov3}. In addition, the initial singularities in the early universe can be absent in such models \cite{Novello1}. Another attractive feature of some NLED is that
there is an upper limit on the electric field at the origin of point-like particles and the self-energy of charges is finite. Thus, in NLED models \cite{Born}-\cite{Kruglov1} there are no problems of singularity and the infinite self-energy of charged particles.
In quantum electrodynamics  nonlinear terms appear due to loop corrections \cite{Heisenberg}-\cite{Adler}.
In this paper we investigate a black hole in the framework of new model of NLED with two parameters.
Black holes also were investigated in the framework of NLED in \cite{Bardeen}-\cite{Frolov} and in many other papers.
In the model under consideration the correspondence principle holds so that
in the weak field limit NLED is converted into Maxwell's electrodynamics.

The paper is organised as follows. In section 2 we propose the model of NLED with two parameters $\beta$ and $\gamma$. In section 3 NLED coupled with general relativity is  formulated.
In sections 4 and 5 the asymptotic of the metric and mass functions at $r\rightarrow 0$ and $r\rightarrow\infty$ are obtained for the parameters $\gamma=1/2$ and $\gamma =3/4$, respectively. We found corrections to the Reissner-Nordstr\"{o}m (RN) solution. The thermodynamics of black holes is investigated. We calculate the Hawking temperature and heat capacity of black holes and demonstrate that  black holes undergo phase transitions of second-order. We obtain the range where black holes are stable. Section 6 is devoted to a conclusion.

The units with $c=\hbar=1$, $\varepsilon_0=\mu_0=1$ are used and the metric signature is $\eta=\mbox{diag}(-1,1,1,1)$.

\section{A model of NLED}

Let us consider NLED with the Lagrangian density
\begin{equation}
{\cal L} = -\frac{{\cal F}}{1+(\beta{\cal F})^\gamma},
 \label{1}
\end{equation}
where the parameter $\beta$ has the dimensions of (length)$^4$, and $\gamma$ is the dimensionless parameter, ${\cal F}=(1/4)F_{\mu\nu}F^{\mu\nu}=(\textbf{B}^2-\textbf{E}^2)/2$,
$F_{\mu\nu}=\partial_\mu A_\nu-\partial_\nu A_\mu$ is the field strength tensor.
At $\gamma=1$ the model was investigated in \cite{Kruglov1}.
At the weak field limit, $\beta {\cal F}\ll 1$, the model (1) becomes Maxwell'l electrodynamics, ${\cal L}\rightarrow-{\cal F}$, i.e. the correspondence principle occurs.

 Making use of Euler-Lagrange equations, from Eq. (1), we obtain field equations
\begin{equation}
\partial_\mu\left({\cal L}_{\cal F}F^{\mu\nu} \right)=0,
\label{2}
\end{equation}
where ${\cal L}_{\cal F}=\partial {\cal L}/\partial{\cal F}$.
From Eq. (1) we find
\begin{equation}\label{3}
  {\cal L}_{\cal F}=\frac{(\gamma-1)(\beta{\cal F})^\gamma-1}{(1+(\beta{\cal F})^\gamma)^2}.
\end{equation}
The electric displacement field $\textbf{D}=\partial{\cal L}/\partial \textbf{E}$, obtained from Eqs. (1) and (3), is given by
\begin{equation}
\textbf{D}=\varepsilon\textbf{E},~~~~\varepsilon=\frac{1-(\gamma-1)(\beta{\cal F})^\gamma}{(1+(\beta{\cal F})^\gamma)^2}.
\label{4}
\end{equation}
The magnetic field $\textbf{H}=-\partial{\cal L}/\partial \textbf{B}$ becomes
\begin{equation}
\textbf{H}=\mu^{-1}\textbf{B},~~~~\mu^{-1}=\varepsilon.
\label{5}
\end{equation}
The field equations (2), making use of Eqs. (4) and (5), may be written as nonlinear Maxwell's equations
\begin{equation}
\nabla\cdot \textbf{D}= 0,~~~~ \frac{\partial\textbf{D}}{\partial
t}-\nabla\times\textbf{H}=0.
\label{6}
\end{equation}
The Bianchi identity, $\partial_\mu \tilde{F}^{\mu\nu}=0$, gives the second pair of nonlinear Maxwell's equations
\begin{equation}
\nabla\cdot \textbf{B}= 0,~~~~ \frac{\partial\textbf{B}}{\partial
t}+\nabla\times\textbf{E}=0.
\label{7}
\end{equation}
From Eqs. (4) and (5), one obtains
\begin{equation}
\textbf{D}\cdot\textbf{H}=(\varepsilon^2)\textbf{E}\cdot\textbf{B}.
\label{8}
\end{equation}
Because $\textbf{D}\cdot\textbf{H}\neq\textbf{E}\cdot\textbf{B}$ \cite{Gibbons} the dual symmetry is broken.
In classical electrodynamics and in Born-Infeld electrodynamics the dual symmetry holds but in QED with loop corrections the dual symmetry is violated.

The causality principle guarantees that the group velocity of excitations over the background is less than the speed of light and requires $ {\cal L}_{\cal F}\leq 0$ \cite{Shabad2}. In this case tachyons will not appear. From Eq. (3) we obtain that at $\gamma\leq 1$ the causality principle holds. In this paper we imply that
$0\leq\gamma\leq 1$ and we consider only magnetized black holes (\textbf{E}=0, ${\cal F}=\textbf{B}^2/2$).

The symmetrical energy-momentum tensor, obtained from Eq. (1), is given by
\begin{equation}
T_{\mu\nu}=\frac{(\gamma-1)(\beta{\cal F})^\gamma-1}{[1+(\beta{\cal F})^\gamma]^2} F_\mu^{~\alpha}F_{\nu\alpha}
-g_{\mu\nu}{\cal L}.
\label{9}
\end{equation}
The trace of the energy-momentum tensor (9) reads
\begin{equation}\label{10}
  {\cal T}\equiv T^{`\mu}_\mu=\frac{4\gamma {\cal F}(\beta{\cal F})^\gamma}{[1+(\beta{\cal F})^\gamma]^2}.
\end{equation}
The trace (10) is not zero because of dimensional parameter $\beta$ so that the scale invariance is broken.

\section{Magnetized black holes}

We will study magnetically charged black hole ($\textbf{E}=0$). The action of NLED coupled with general relativity is
\begin{equation}
I=\int d^4x\sqrt{-g}\left(\frac{1}{2\kappa^2}R+ {\cal L}\right),
\label{11}
\end{equation}
where $\kappa^2=8\pi G\equiv M_{Pl}^{-2}$, $G$ is Newton's constant, $M_{Pl}$ is the reduced Planck mass, and $R$ is the Ricci scalar.
By varying action (11) with respect of the metric and electric potential, we obtain the Einstein  and electromagnetic field equations
\begin{equation}
R_{\mu\nu}-\frac{1}{2}g_{\mu\nu}R=-\kappa^2T_{\mu\nu},
\label{12}
\end{equation}
\begin{equation}
\partial_\mu\left[\sqrt{-g}{\cal L}_{\cal F}F^{\mu\nu}\right]=0.
\label{13}
\end{equation}
We consider the line element having the spherical symmetry
\begin{equation}
ds^2=-f(r)dt^2+\frac{1}{f(r)}dr^2+r^2(d\vartheta^2+\sin^2\vartheta d\phi^2).
\label{14}
\end{equation}
The metric function is defined by the relation \cite{Bronnikov}
\begin{equation}
f(r)=1-\frac{2GM(r)}{r},
\label{15}
\end{equation}
and the mass function is given by
\begin{equation}
M(r)=\int_0^r\rho_M(r)r^2dr,
\label{16}
\end{equation}
where $\rho_M$ is the magnetic energy density. The $m_M=\int_0^\infty\rho_M(r)r^2dr$ is the magnetic mass of the black hole possessing the electromagnetic nature.
The magnetic energy density (\textbf{E}=0), found from Eq. (9), is
\begin{equation}\label{17}
  \rho_M=T_0^{~0}=\frac{{\cal F}}{1+(\beta{\cal F})^\gamma}.
\end{equation}
We have for the magnetized black hole \cite{Bronnikov} ${\cal F}=q^2/(2r^4)$, where $q$ is a magnetic charge. For arbitrary value of $\gamma$ the expression for $M(r)$ is complicated. Therefore, we investigate the particular cases $\gamma=1/2$ and $\gamma=3/4$.

\section{The case $\gamma=\frac{1}{2}$}

Introducing the dimensional parameter $x=2^{1/4}r/(\beta^{1/4}\sqrt{q})$ and
making use of Eq. (17) we obtain the magnetic energy density
\begin{equation}\label{18}
  \rho_M(x)=\frac{1}{\beta x^2(1+x^2)}.
\end{equation}
From Eqs. (16) and (18) one finds the mass function
\begin{equation}\label{19}
  M(x)=\frac{q^{3/2}}{2^{3/4}\beta^{1/4}}\arctan (x).
\end{equation}
From Eq. (19) we obtain the magnetic mass of the black hole
\begin{equation}\label{20}
 m_M = M(\infty)=\frac{\pi q^{3/2}}{2^{7/4}\beta^{1/4}}\simeq \frac{0.934 q^{3/2}}{\beta^{1/4}}.
\end{equation}
Making use of Eqs. (15) and (19) we obtain the metric function expressed through the dimensionless variable
\begin{equation}\label{21}
  f(x)=1-\frac{\sqrt{2}Gq}{\sqrt{\beta}x}\arctan x.
\end{equation}
From Eq. (21) one finds the asymptotic of the metric function at $r\rightarrow 0$ and $r\rightarrow\infty$
\begin{equation}\label{22}
 f(r)=1-\frac{\sqrt{2}Gq}{\sqrt{\beta}} +\frac{2Gr^2}{\beta}-\frac{2^{3/2}Gr^4}{\beta^{3/2}q}+{\cal O}(r^6)~~~~r\rightarrow 0,
\end{equation}
\begin{equation}\label{23}
  f(r)=1-\frac{2Gm_M}{r}+\frac{Gq^2}{r^2}-\frac{G\sqrt{\beta}q^3}{3\sqrt{2} r^4}+
{\cal O}(r^{-6})~~~~r\rightarrow \infty.
\end{equation}
It should me mentioned that according to Eq. (22) the black hole does not have the regular solution because $f(0)\neq 1$.
Equation (23) gives corrections to RN solution in the order of ${\cal O}(r^{-4})$.
At $r\rightarrow \infty$ we have $f(\infty)=1$ and the spacetime becomes flat. If $\beta=0$ our model becomes Maxwell's electrodynamics and Eq. (23) is converted into the RN solution.
The plots of the function $f(x)$ for different parameters $a=\sqrt{\beta}/(\sqrt{2}Gq)$ are represented in Fig. 1.
\begin{figure}[h]
\includegraphics[height=3.0in,width=3.0in]{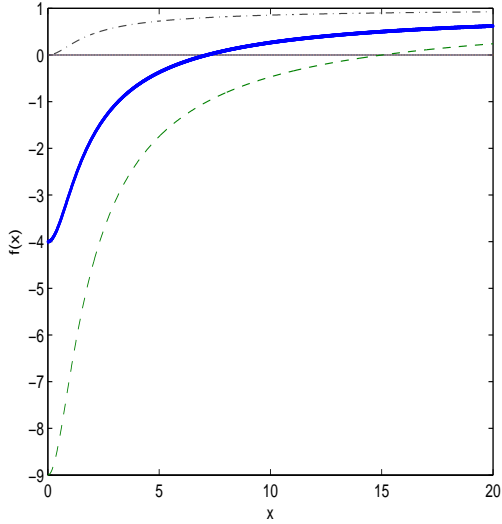}
\caption{\label{fig.1}The plot of the function $f(x)$. Dashed-dotted line corresponds to $a=1$, solid line corresponds to $a=0.2$ and dashed line corresponds to $a=0.1$.}
\end{figure}
Fig. 1 shows that black holes may have one or no horizons depending on the parameter $a$.
The horizons of the black hole are defined by equation $f(x_+)=0$ and
represented in Table 1.
\begin{table}[ht]
\caption{The horizons of the black hole}
\centering
\begin{tabular}{c c c c c c c c c c }\\[1ex]
\hline \hline
$a$ & 0.1 & 0.2 & 0.3 & 0.4 & 0.5 & 0.6 & 0.7 & 0.8 & 0.9  \\[0.5ex]
\hline
$x_+$ & 15.044 & 7.160 & 4.508 & 3.161 & 2.331 & 1.755 & 1.315 & 0.949 & 0.603 \\[0.5ex]
\hline
\end{tabular}
\end{table}

The Ricci scalar found from Eqs. (10) and (12) (for $\gamma=0.5$, $\textbf{E}=0$) is given by
\begin{equation}
R=\kappa^2 {\cal T}=\kappa^2\frac{\sqrt{2\beta}q^3}{r^2(\sqrt{2}r^2+\sqrt{\beta}q)^2}.
\label{24}
\end{equation}
At $r\rightarrow \infty$  the Ricci scalar (24) goes to zero, $R\rightarrow 0$, and spacetime becomes flat.

\subsection{Thermodynamics}

Let us study the thermal stability of magnetically charged black holes. We use the expression for the Hawking temperature
\begin{equation}
T_H=\frac{\kappa_S}{2\pi}=\frac{f'(r_+)}{4\pi}.
\label{25}
\end{equation}
Here $\kappa_S$ is the surface gravity and $r_+$ is the event horizon. From Eqs. (15) and (16) one obtains the relations as follows:
\begin{equation}
f'(r)=\frac{2 GM(r)}{r^2}-\frac{2GM'(r)}{r},~~~M'(r)=r^2\rho_M,~~~M(r_+)=\frac{r_+}{2G}.
\label{26}
\end{equation}
Making use of Eqs. (21), (25) and (26) one finds the Hawking temperature
\begin{equation}
T_H=\frac{1}{2^{7/4}\pi\beta^{1/4}\sqrt{q}} \left(\frac{1}{x_+}-\frac{1}{(1+x^2_+)\arctan(x_+)}\right).
\label{27}
\end{equation}
The plot of the Hawking temperature versus the horizon $x_+$ is represented in Fig. 2.
\begin{figure}[h]
\includegraphics[height=3.0in,width=3.0in]{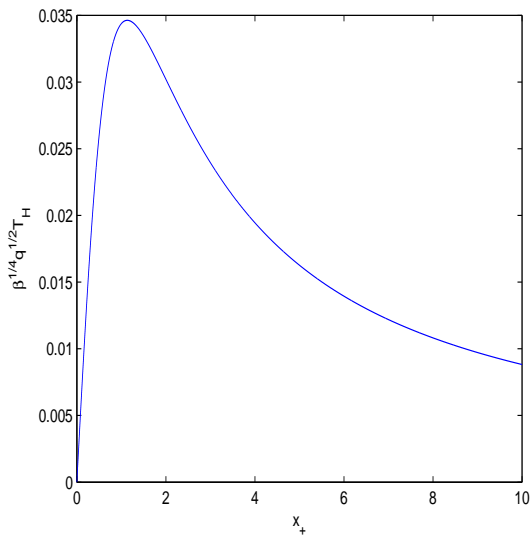}
\caption{\label{fig.2}The plot of the function $T_H\sqrt{q}\beta^{1/4}$ vs. $x_+$}.
\end{figure}
It follows from Fig. 2 that the Hawking temperature is positive, and therefore, the black hole is stable. The first-order phase transition occurs when the Hawking temperature changes the sign. The second-order phase transition takes place if the heat capacity is singular \cite{Davies}.
We use the entropy which satisfies the Hawking area low $S=A/(4G)=\pi r_+^2/G$.
Than the heat capacity can be calculated by
\begin{equation}
C_q=T_H\left(\frac{\partial S}{\partial T_H}\right)_q=\frac{T_H\partial S/\partial r_+}{\partial T_H/\partial r_+}=\frac{2\pi r_+T_H}{G\partial T_H/\partial r_+}.
\label{28}
\end{equation}
According to Eq. (28) the heat capacity diverges if the Hawking temperature has extremum.
The maximum of the Hawking temperature occurs at $x_+\simeq 1.135$, and in this point there is the phase transition of the second-order. From Eqs. (27) and (28) we obtain the heat capacity
\begin{equation}\label{29}
  C_q=\frac{\pi\sqrt{2\beta}q[(1+x_+^2)\arctan(x_+)-x_+]x_+^2(1+x_+^2)\arctan(x_+)}
{G[x_+^2(1+2x_+\arctan(x_+))-(1+x_+^2)^2\arctan^2(x_+)]}.
\end{equation}
The plot of the function $GC_q/(\sqrt{\beta}q)$ vs. $ x_+$ is given in Fig. 3.
\begin{figure}[h]
\includegraphics[height=3.0in,width=3.0in]{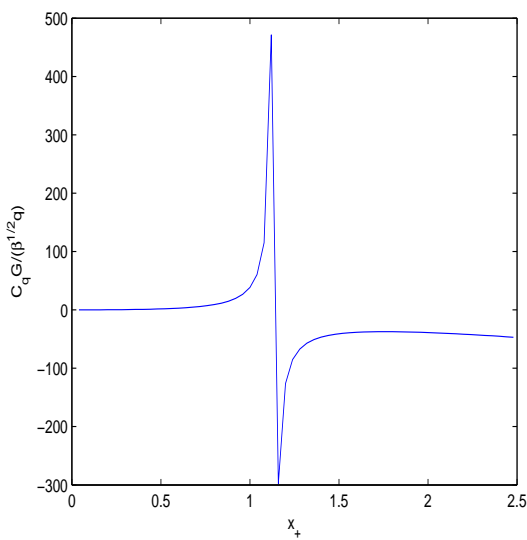}
\caption{\label{fig.3}The plot of the function $C_qG/(\sqrt{\beta}q)$ vs. $x_+$.}
\end{figure}
Fig. 3 shows that indeed at $x_+\simeq 1.135$ the black hole undergoes the second-order phase transition. At $0\leq x_+<1.135$ the black hole is stable and at $x_+>1.135$ the heat capacity is negative and
the black hole becomes unstable.

\section{The case $\gamma=\frac{3}{4}$}

With the help of Eq. (17) and using the dimensional parameter $x=2^{1/4}r/(\beta^{1/4}\sqrt{q})$, one finds the magnetic energy density
\begin{equation}\label{30}
  \rho_M(x)=\frac{1}{\beta x(1+x^3)}.
\end{equation}
From Eqs. (16) and (30) we obtain the mass function
\begin{equation}\label{31}
  M(x)=\frac{q^{3/2}}{2^{7/4}3\beta^{1/4}}\left[\ln\frac{x^2-x+1}{(x+1)^2}+2\sqrt{3} \arctan \left(\frac{2x-1}{\sqrt{3}}\right)\right].
\end{equation}
From Eq. (31) one finds the magnetic mass of the black hole
\begin{equation}\label{32}
 m_M= M(\infty)=\frac{\pi q^{3/2}}{2^{7/4}\sqrt{3}\beta^{1/4}}\simeq \frac{0.54 q^{3/2}}{\beta^{1/4}}.
\end{equation}
By virtue of Eqs. (15) and (31) we obtain the metric function
\begin{equation}\label{33}
  f(x)=1-\frac{Gq}{3\sqrt{2\beta}x}\left[\ln\frac{x^2-x+1}{(x+1)^2}+2\sqrt{3} \arctan \left(\frac{2x-1}{\sqrt{3}}\right)\right].
\end{equation}
Making use of Eq. (33) one finds the asymptotic of the metric function at $r\rightarrow 0$ and $r\rightarrow\infty$
\begin{equation}\label{34}
 f(r)=1+\frac{\pi Gq^{3/2}}{3\sqrt{6}\beta^{1/4}r} -\frac{G\sqrt{q}r}{2^{1/4}\beta^{3/4}}+\frac{2^{3/2}Gr^4}{5\beta^{3/2}q}+{\cal O}(r^6)~~~~r\rightarrow 0,
\end{equation}
\begin{equation}\label{35}
  f(r)=1-\frac{2Gm_M}{r}+\frac{Gq^2}{r^2}-\frac{G\beta^{3/4}q^{7/2}}{2^{11/4} r^5}+
{\cal O}(r^{-6})~~~~r\rightarrow \infty.
\end{equation}
It follows from Eq. (34) that $f(0)= \infty$, and therefore, the black hole solution is not the regular solution.
In accordance with Eq. (35) corrections to RN solution are in the order of ${\cal O}(r^{-5})$.
When $r\rightarrow \infty$ the spacetime becomes Minkowski's spacetime. At $\beta=0$ the model is transformed into Maxwell's electrodynamics and Eq. (35) becomes the RN solution.
The plots of the function $f(x)$ for different parameters $a=\sqrt{\beta}/(\sqrt{2}Gq)$ are represented in Fig. 4.
\begin{figure}[h]
\includegraphics[height=3.0in,width=3.0in]{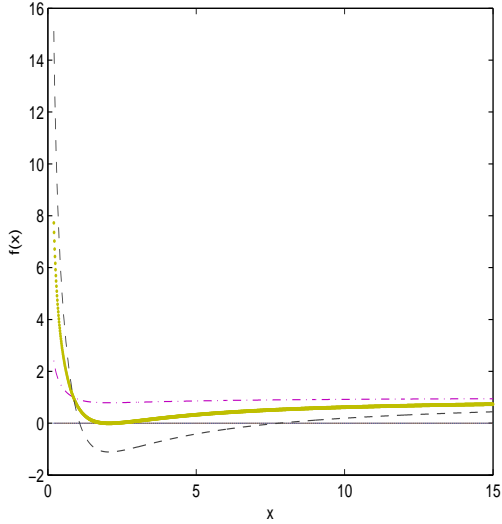}
\caption{\label{fig.4}The plot of the function $f(x)$. Dashed-dotted line corresponds to $a=1$, solid line corresponds to $a=0.21$ and dashed line corresponds to $a=0.1$.}
\end{figure}
Fig. 4 shows different behaviour of the function $f(x)$ depending on the parameter $a$. At $a>0.21$ there are no horizons and we have naked singularity. When $a\simeq 0.21$ the extreme singularity occurs. At $a<0.21$ one has two horizons.
The inner $x_-$ and outer $x_+$ horizons of the black hole are given by the equation $f(x)=0$ and
are in Table 2.
\begin{table}[ht]
\caption{The horizons of the black hole}
\centering
\begin{tabular}{c c c c c c c c c c }\\[1ex]
\hline \hline
$a$ & 0.05 & 0.08 & 0.1 & 0.13 & 0.15 & 0.18 & 0.2 & 0.210 & 0.211  \\[0.5ex]
\hline
$x_-$ & 0.953 & 1.021 & 1.074 & 1.170 & 1.252 & 1.427 & 1.642 & 1.923 & 2.039 \\[0.5ex]
\hline
$x_+$ & 16.959 & 10.099 & 7.785 & 5.606 & 4.601 & 3.427 & 2.716 & 2.222 & 2.088 \\[0.5ex]
\hline
\end{tabular}
\end{table}

The Ricci scalar due to Eqs. (10) and (12), for $\gamma=3/4$ ($\textbf{E}=0$)
at $r\rightarrow \infty$, approaches to zero, $R\rightarrow 0$, and spacetime becomes Minkowski's spacetime.

\subsection{Thermodynamics}

By virtue of Eqs. (25), (26) and (30) we obtain the Hawking temperature
\begin{equation}
T_H=\frac{1}{2^{7/4}\pi\beta^{1/4}\sqrt{q}} \left(\frac{1}{x_+}-\frac{6x_+}{(x_+^3+1)\left[\ln\frac{x^2_+-x_++1}{(x_++1)^2}+2\sqrt{3}
\arctan\left(\frac{2x_+-1}{\sqrt{3}}\right)\right]}\right).
\label{36}
\end{equation}
The plot of the function$T_H(x_+)$ is given in Fig. 5.
\begin{figure}[h]
\includegraphics[height=3.0in,width=3.0in]{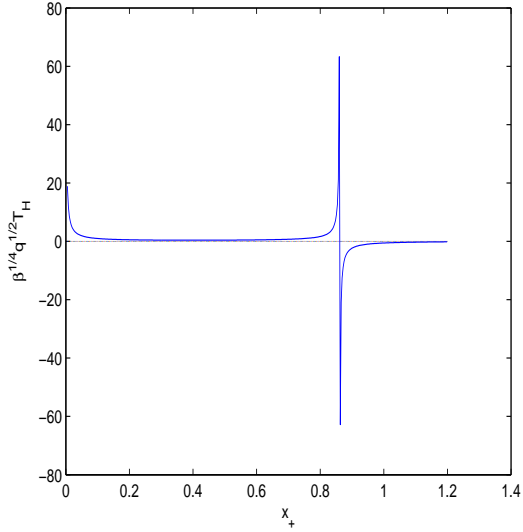}
\caption{\label{fig.5}The plot of the function $T_H\sqrt{q}\beta^{1/4}$ vs. $x_+$}.
\end{figure}
According to Fig. 5 the Hawking temperature is positive at $0.862>x_+>0$, and
the black hole is unstable for $x_+>0.862$. It follows from Fig. 5 that the Hawking temperature possesses the minimum at $x_+\simeq 0.4$ where
the second-order phase transition takes place.  In this point the heat capacity is singular (see Eq. (28)).
From Eq. (36) we obtain the derivative of the Hawking temperature
\[
\frac{\partial T_H}{\partial x_+}=\frac{1}{2^{7/4}\pi\beta^{1/4}\sqrt{q}} \biggl[-\frac{1}{x_+^2}
\]
\begin{equation}
+\frac{6\left((2x_+^3-1)\ln\left(\frac{x_+^2-x_++1}{(x_++1)^2}\right)+2\sqrt{3}(2x_+^3-1)\arctan\left(\frac{2x_+-1}
{\sqrt{3}}\right)+6x_+^2\right)}{(x_+^3+1)^2\left[\ln\left(\frac{x_+^2-x_++1}{(x_++1)^2}\right)+2\sqrt{3}
\arctan\left(\frac{2x_+-1}{\sqrt{3}}\right)\right]^2}\biggr].
\label{37}
\end{equation}
From Eqs. (28), (36) and (37) one can find the heat capacity.
The plot of the function $GC_q/(\sqrt{\beta}q)$ vs. $ x_+$ is given in Fig. 6.
\begin{figure}[h]
\includegraphics[height=3.0in,width=3.0in]{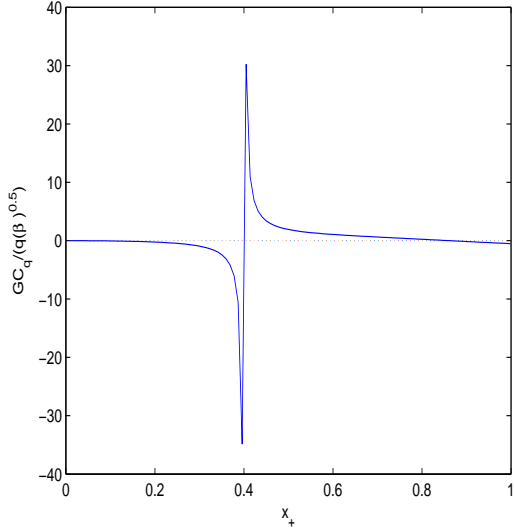}
\caption{\label{fig.6}The plot of the function $C_qG/(\sqrt{\beta}q)$ vs. $x_+$.}
\end{figure}
Fig. 6 shows that at $x_+\simeq 0.4$ the black hole has the second-order phase transition. When $0< x_+<0.4$ the heat capacity is negative and
the black hole is unstable but at $0.862>x_+>0.4$ the black hole is stable.

 \section{Conclusion}

We have proposed new NLED model with two independent parameters $\beta$ and $\gamma$. In this model the correspondence principle takes place so that for weak fields the model is converted into Maxwell's electrodynamics. NLED coupled with the gravitational field was investigated.
We have studied the magnetized black holes and found the asymptotic of the metric and mass functions at $r\rightarrow 0$ and $r\rightarrow\infty$ for the parameters $\gamma=1/2$ and $\gamma =3/4$.
Corrections to the RN solution were found. We have calculated the Hawking temperature and heat capacity  of black holes and demonstrated that  black holes undergo second-order phase transitions.
 The range of horizons where black holes are stable was obtained.

In this paper we have studied only the thermodynamic stabilities of black holes. It is of
interest to investigate the stability of black holes, within our model, against
perturbations. In this case one can compare the stability of our black holes with the
Reissner-Nordstr\"{o}m black hole. We leave this for further investigations.

\end{document}